# THE STANDARD MODEL AND FERMION MASSES

V.P. Neznamov

RFNC-VNIIEF, 607190 Sarov, Nizhniy Novgorod region
e-mail: Neznamov@vniief.ru

Abstract

The paper formulates the Standard Model with massive fermions without introduction of the Yukawa interaction of Higgs bosons with fermions.

With such approach, Higgs bosons are responsible only for the gauge invariance of the theory's boson sector and interact only with gauge bosons $W_\mu^\pm$, $Z_\mu$, gluons, and photons.

The papers [1], [2] offer the Standard Model in the modified Foldy-Wouthuysen representation. It has been shown that for the being *SU(2)*-invariant, the theory formulated in the Foldy-Wouthuysen representation does not necessarily require Higgs bosons to interact with fermions, while all theoretical and experimental implications of the Standard Model obtained in the Dirac representation are preserved. The goal of the present paper is to give a similar formulation of the Standard Model in the Dirac representation with massive fermions meeting the requirements of local *SU(3)×SU(2)×U(1)* symmetry.

The paper uses the system of units with $\hbar = c = 1$; $x, p, B$ are 4-vectors; the inner product is taken in the form

$$xy = x^\mu y_\mu = x^0 y^0 - x^k y^k, \mu = 0,1,2,3; k = 1,2,3; p^\mu = i\frac{\partial}{\partial x_\mu}; \partial^\mu = \frac{\partial}{\partial x_\mu};$$

$$\alpha^\mu = \begin{cases} 1, \mu = 0 \\ \alpha^k, \mu = k = 1,2,3 \end{cases}; \gamma^\mu = \gamma^0 \alpha^\mu; \beta = \gamma^0, \alpha^k, \gamma^\mu, \gamma^5 - \text{Dirac matrices}$$

Consider the density of Hamiltonian of a Dirac particle with mass $m_f$ interacting with an arbitrary boson field $B^\mu$

$$\mathcal{H}_D = \psi^\dagger (\vec{\alpha}\vec{p} + \beta m_f + q\alpha_\mu B^\mu)\psi = \psi^\dagger (P_L + P_R)(\vec{\alpha}\vec{p} + \beta m_f + q\alpha_\mu B^\mu)(P_L + P_R)\psi =$$

$$= \psi_L^\dagger (\vec{\alpha}\vec{p} + q\alpha_\mu B^\mu)\psi_L + \psi_R^\dagger (\vec{\alpha}\vec{p} + q\alpha_\mu B^\mu)\psi_R + \psi_L^\dagger \beta m_f \psi_R + \psi_R^\dagger \beta m_f \psi_L \quad . \quad (1)$$

In (1) $q$ – is the coupling constant; $P_L = \frac{1-\gamma_5}{2}, P_R = \frac{1+\gamma_5}{2}$ – are the left and right projective operators; $\psi_L = P_L \psi, \psi_R = P_R \cdot \psi$ – are the left and right components of the operator of the Dirac field $\psi$.

The density of Hamiltonian $\mathcal{H}_D$ allows obtaining the motion equations for $\psi_L$ and $\psi_R$:

$$p_0 \psi_L = (\vec{\alpha}\vec{p} + q\alpha_\mu B^\mu)\psi_L + \beta m_f \psi_R$$
$$p_0 \psi_R = (\vec{\alpha}\vec{p} + q\alpha_\mu B^\mu)\psi_R + \beta m_f \psi_L \quad . \quad (2)$$

One can see that both the density of Hamiltonian $\mathcal{H}_D$ and motion equations have a form, which is not $SU(2)$-invariant because of the Dirac fermion having a mass. For this reason we first consider massless fermions in the Standard Model to provide $SU(2)$–invariance and then Dirac fermions are given masses after the mechanism of spontaneous violation of the symmetry has been introduced and Higgs bosons have appeared. [3]

There is a question: Is it possible to find the form of Hamiltonian and the forms of motion equations for massive fermions with their $SU(2)$–symmetry preserved?

If follows from equations (2) that

$$\psi_L = \left(p_0 - \vec{\alpha}\vec{p} - q\alpha_\mu B^\mu\right)^{-1} \beta m_f \psi_R$$
$$\psi_R = \left(p_0 - \vec{\alpha}\vec{p} - q\alpha_\mu B^\mu\right)^{-1} \beta m_f \psi_R \qquad (3)$$

Substitute (3) into the right-hand parts of equations (2) proportional to $\beta m_f$ and obtain the integro-differential equations for $\psi_R$ and $\psi_L$:

$$\left[\left(p_0 - \vec{\alpha}\vec{p} - q\left(\alpha_0 B^0 - \vec{\alpha}\vec{B}\right)\right) - \beta m_f \left(p_0 - \vec{\alpha}\vec{p} - q\left(\alpha_0 B^0 - \vec{\alpha}\vec{B}\right)\right)^{-1} \beta m_f\right]\psi_L = 0$$
$$\left[\left(p_0 - \vec{\alpha}\vec{p} - q\left(\alpha_0 B^0 - \vec{\alpha}\vec{B}\right)\right) - \beta m_f \left(p_0 - \vec{\alpha}\vec{p} - q\left(\alpha_0 B^0 - \vec{\alpha}\vec{B}\right)\right)^{-1} \beta m_f\right]\psi_R = 0 \qquad (4)$$

It is clearly seen that equations for $\psi_R$ and $\psi_L$ have the same form and, in contrast to equations (2), the presence of mass $m_f$ does not lead to mixing the left and right components of $\psi$.

Equations (4) can be written as

$$\left[\left(p_0 - \vec{\alpha}\vec{p} - q\alpha_\mu B^\mu\right) - \left(p_0 + \vec{\alpha}\vec{p} - q\bar{\alpha}_\mu B^\mu\right)^{-1} m^2\right]\psi_{L,R} = 0. \qquad (5)$$

In expression (5), $\psi_{L,R}$ shows that equations for $\psi_L$ and $\psi_R$ have the same form;

$$\bar{\alpha}_\mu = \begin{cases} 1 \\ -\alpha^i \end{cases}.$$

If we multiply (on the left side) equations (5) by term $p_0 + \vec{\alpha}\vec{p} - q\bar{\alpha}_\mu B^\mu$, we obtain the second-order equations with respect to $p^\mu$:

$$\left[\left(p_0 + \vec{\alpha}\vec{p} - q\bar{\alpha}_\mu B^\mu\right)\left(p_0 - \vec{\alpha}\vec{p} - q\alpha_\mu B^\mu\right) - m^2\right]\psi_{L,R} = 0. \qquad (6)$$

For quantum electrodynamics $\left(q = e, B^\mu = A^\mu\right)$ equations (6) have the form

$$\left[\left(p_0 - eA_0\right)^2 - \left(\vec{p} - e\vec{A}\right)^2 - m^2 + e\vec{\sigma}\vec{H} + i\vec{\alpha}\vec{E}\right]\psi_{L,R} = 0. \qquad (7)$$

In equation (7), $\vec{H} = rot\,\vec{A}$ – is magnetic field, $\vec{E} = -\dfrac{\partial \vec{A}}{\partial t} - \nabla A_0$ – is electrical field.

Equations (7) coincide with the second-order equation obtained by Dirac in the 1920s [4]. However, in contrast to [4] (see also [5]), equations (7) contain no «excess» solutions. The operator $\gamma_5$ commutes with equations (6). Consequently, $\gamma_5 \psi = \delta \psi$ $(\delta^2 = 1; \delta = \pm 1)$. The case of $\gamma_5 = -1$ corresponds to the solution of equation (7) for $\psi_L$ and $\gamma_5 = +1$ corresponds to the solution of equation (7) for $\psi_R$.

Equations (5), (6) are invariant relative to $SU(2)$ – transformations, however, they are nonlinear relative to operator $p_0 = i\dfrac{\partial}{\partial t}$. Linear forms of $SU(2)$ – invariant equations for fermion fields relative to $p_0$ can be obtained using the Foldy-Wouthuysen transformation [6] in the specially introduced isotopic space.

Introduce the eight-component field operator $\Phi_1 = \begin{pmatrix} \psi_L \\ \psi_R \end{pmatrix}$ and isotopic matrices $\tau_3 = \begin{pmatrix} I & 0 \\ 0 & -I \end{pmatrix}, \tau_1 = \begin{pmatrix} 0 & I \\ I & 0 \end{pmatrix}$ affecting the four upper and four lower components of operator $\Phi_1$. Now, equations (2) can be written as

$$p_0 \Phi_1 = \left(\vec{\alpha}\vec{p} + \tau_1 \beta m_f + q\alpha_\mu B^\mu\right)\Phi_1. \tag{8}$$

Owing to commutation between $\tau_1$ and the right-hand part of equation (8), field $\Phi_2 = \tau_1 \Phi_1 = \begin{pmatrix} \psi_R \\ \psi_L \end{pmatrix}$ is also solution to equation (8).

Further, consider equation (8) without boson field $B^\mu$ (free motion):

$$p_0 \Phi_{1,2} = \left(\vec{\alpha}\vec{p} + \tau_1 \beta m_f\right)\Phi_{1,2}. \tag{9}$$

$\Phi_{1,2}$ is written to show that equations (9) are the same for operators $\Phi_1$, $\Phi_2$.

Find the Foldy-Wouthuysen transformation in isotopic space for the free motion equation (9) using the Eriksen transformation [7].

$$U^0_{FW} = U_{Er} = \frac{1}{2}(1+\tau_3 \lambda)\left(\frac{1}{2} + \frac{\tau_3 \lambda + \lambda \tau_3}{4}\right)^{-\frac{1}{2}} \tag{10}$$

In expression (10) we have $\lambda = \dfrac{\vec{\alpha}\vec{p} + \tau_1 \beta m_f}{E}$; $E = (\vec{p}^2 + m^2)^{1/2}$. Since we have $(\vec{\alpha}\vec{p} + \tau_1 \beta m_f)^2 = E^2$, then $\lambda^2 = 1$.

Expression (10) can be transformed to obtain the following expression:

$$U^0_{FW} = U_{Er} = \frac{1}{2}\left(1 + \frac{\tau_3 \vec{\alpha}\vec{p} + \tau_3 \tau_1 \beta m}{E}\right)\left(\frac{1}{2} + \frac{\tau_3 \vec{\alpha}\vec{p}}{2E}\right)^{-1/2} = $$
$$= \sqrt{\frac{E + \tau_3 \vec{\alpha}\vec{p}}{2E}}\left(1 + \frac{1}{E + \tau_3 \vec{\alpha}\vec{p}}\tau_3 \tau_1 \beta m\right) \tag{11}$$

Transformation (11) is a unitary transformation $\left(U^0_{FW}(U^0_{FW})^\dagger = 1\right)$ and

$$U^0_{FW}(\vec{\alpha}\vec{p} + \tau_1 \beta m_f)(U^0_{FW})^\dagger = \tau_3 E \tag{12}$$

Thus, equations (9) in the Foldy-Wouthuysen representation have the form

$$p_0 (\Phi_{1,2})_{FW} = \tau_3 E (\Phi_{1,2})_{FW}. \tag{13}$$

Now, let us check whether the sufficiency condition for transformation to the Foldy-Wouthuysen representation [8] is fulfilled, or not.

With regard to relations (2), (3) the normalized solutions to equation (9) for field $\Phi_1$ can be written, as follows:

Positive-energy solution

$$\Phi_1^{(+)} = e^{-iEt}\sqrt{\frac{E + \tau_3 \vec{\alpha}\vec{p}}{2E}}\begin{pmatrix} \psi_R \\ \dfrac{1}{E - \vec{\alpha}\vec{p}}\beta m \psi_R \end{pmatrix} = e^{-iEt}\begin{pmatrix} \sqrt{\dfrac{E + \vec{\alpha}\vec{p}}{2E}}\psi_R \\ \dfrac{1}{\sqrt{2E(E - \vec{\alpha}\vec{p})}}\beta m \psi_R \end{pmatrix} \tag{14}$$

Negative-energy solution

$$\Phi_1^{(-)} = e^{iEt}\sqrt{\frac{E + \tau_3 \vec{\alpha}\vec{p}}{2E}}\begin{pmatrix} -\dfrac{1}{E + \vec{\alpha}\vec{p}}\beta m \psi_L \\ \psi_L \end{pmatrix} = e^{iEt}\begin{pmatrix} -\dfrac{1}{\sqrt{2E(E + \vec{\alpha}\vec{p})}}\beta m \psi_L \\ \sqrt{\dfrac{E - \vec{\alpha}\vec{p}}{2E}}\psi_L \end{pmatrix} \tag{15}$$

By applying matrix $U^0_{FW}$ to $\Phi_1^{(+)}, \Phi_1^{(-)}$ we obtain

$$\left(\Phi_1^{(+)}\right)_{FW} = U^0_{FW}\Phi_1^{(+)} = e^{-iEt}\begin{pmatrix} \psi_R^{(+)} \\ 0 \end{pmatrix}$$
$$\left(\Phi_1^{(-)}\right)_{FW} = U^0_{FW}\Phi_1^{(-)} = e^{iEt}\begin{pmatrix} 0 \\ \psi_L^{(-)} \end{pmatrix}. \tag{16}$$

One can see from (16) that the sufficiency condition is fulfilled and matrix $U_{FW}^0$ is, indeed, the Foldy-Wouthuysen transformation for spinor $\Phi_1$ in the isotopic space we have introduced.

Similarly, for field $\Phi_2 = \begin{pmatrix} \psi_L \\ \psi_R \end{pmatrix}$ we can obtain

$$\left(\Phi_2^{(+)}\right)_{FW} = U_{FW}^0 \Phi_2^{(+)} = e^{-iEt} \begin{pmatrix} \psi_L^{(+)} \\ 0 \end{pmatrix}$$
$$\left(\Phi_2^{(-)}\right)_{FW} = U_{FW}^0 \Phi_2^{(-)} = e^{iEt} \begin{pmatrix} 0 \\ \psi_R^{(-)} \end{pmatrix} \tag{17}$$

Equations (13) allow writing the density of the free motion Hamiltonian for fermions of mass $m_f$ in the form

$$\mathcal{H}_{FW} = (\Phi_1)_{FW}^\dagger \tau_3 E (\Phi_1)_{FW} + (\Phi_2)_{FW}^\dagger \tau_3 E (\Phi_2)_{FW} = (\Phi_1^{(+)})_{FW}^\dagger E (\Phi_1^{(+)})_{FW} - (\Phi_1^{(-)})_{FW}^\dagger E (\Phi_1^{(-)})_{FW} +$$
$$+ (\Phi_2^{(+)})_{FW}^\dagger E (\Phi_2^{(+)})_{FW} - (\Phi_2^{(-)})_{FW}^\dagger E (\Phi_2^{(-)})_{FW} = \left(\psi_R^{(+)}\right)^\dagger E \psi_R^{(+)} - \left(\psi_L^{(-)}\right)^\dagger E \psi_L^{(-)} + \tag{18}$$
$$+ \left(\psi_L^{(+)}\right)^\dagger E \psi_L^{(+)} - \left(\psi_R^{(-)}\right)^\dagger E \psi_R^{(-)}.$$

In the presence of boson fields $B^\mu(x)$ interacting with fermion fields $\Phi_1(x), \Phi_2(x)$, the Foldy-Wouthuysen transformation and Hamiltonian of equation (8) in the Foldy-Wouthuysen representation in isotopic space could be obtained as a series in powers of the coupling constant using the algorithm described in [1], [9].

As a result, using denotations from [1], [9] we obtain

$$U_{FW} = U_{FW}^0 \left(1 + \delta_1 + \delta_2 + \delta_3 + \ldots\right) \tag{19}$$

$$p_0 (\Phi_{1,2})_{FW} = H_{FW} (\Phi_{1,2})_{FW} = \left(\tau_3 E + q K_1 + q^2 K_2 + q^3 K_3 + \ldots\right)(\Phi_{1,2})_{FW}. \tag{20}$$

Expressions for operators $C$ and $N$ constituting the basis for the interaction Hamiltonian in the Foldy-Wouthuysen representation can be written in the following form in our case:

$$C = \left[U_{FW}^0 q \alpha_\mu B^\mu \left(U_{FW}^0\right)^\dagger\right]^{even} = qR\left(B^0 - LB^0 L\right)R - qR\left(\vec{\alpha}\vec{B} - L\vec{\alpha}\vec{B}L\right)R$$
$$N = \left[U_{FW}^0 q \alpha_\mu B^\mu \left(U_{FW}^0\right)^\dagger\right]^{odd} = qR\left(LB^0 - B^0 L\right)R - qR\left(L\vec{\alpha}\vec{B} - \vec{\alpha}\vec{B}L\right)R \tag{21}$$
$$R = \sqrt{\frac{E + \tau_3 \vec{\alpha}\vec{p}}{2E}}; \quad L = \frac{1}{E + \tau_3 \vec{\alpha}\vec{p}} \tau_3 \tau_1 \beta m$$

The superscripts *even* and *odd* in expressions (21) show the even and odd parts of operators relative to the upper and lower components of $\Phi_1$ and $\Phi_2$.

For equations (20) we can write the Hamiltonian density for fermion fields $(\Phi_1)_{FW}, (\Phi_2)_{FW}$, interacting with boson field $B^\mu(x)$.

$$\mathscr{H}_{FW} = (\Phi_1)_{FW}^\dagger \left( \tau_3 E + qK_1 + q^2 K_2 + q^3 K_3 + ... \right) (\Phi_1)_{FW} +$$

$$+ (\Phi_2)_{FW}^\dagger \left( \tau_3 E + qK_1 + q^2 K_2 + q^3 K_3 + ... \right) (\Phi_2)_{FW} \qquad (22)$$

Considering the fact that the expression for the Foldy-Wouthuysen Hamiltonian in parentheses in equation (22) has, by definition, the diagonal form relative to the upper and lower components $(\Phi_1)_{FW}, (\Phi_2)_{FW}$ [6], [8], [9], it is clear that the Hamiltonian density (22) is $SU(2)$–invariant.

Thus, expression (22) and equations (20) are invariant relative to $SU(2)$-transformations independently of the presence or absence of fermion masses.

Expression (22) demonstrates the necessity of using two fermion fields - $(\Phi_1)_{FW}(x), (\Phi_2)_{FW}(x)$ - in the formalism. If only $(\Phi_1)_{FW}(x)$ is used in the theory, motion and interactions of the right fermions, as well as motion and interactions of left antifermions remain; and vice versa, if only $(\Phi_2)_{FW}(x)$ is used in the theory, motion and interaction of the left fermions, as well as motion and interactions of the right antifermions remain.

We can arrive at using the two fermion fields - $(\Phi_1)_{FW}(x), (\Phi_2)_{FW}(x)$ - to describe the motion of a fermion with mass $m_f$, if we agree with the necessity to include the states of equations (2) with the sign changed before the mass term [1]. Changing the sign before mass $m_f$ in Dirac equation, in itself, leads to no physical consequences, however, when changing to the Foldy-Wouthuysen representation, a complete description of the motion and interactions of both the left and right fermions can be obtained using two fermion fields. In this case, Dirac equations for fields $\Phi_1, \Phi_2$ have the form

$$\begin{aligned} p_0 \Phi_1(x) &= \left( \vec{\alpha}\vec{p} + \tau_1 \beta m_f + q\alpha_\mu B^\mu \right) \Phi_1(x) \\ p_0 \Phi_2(x) &= \left( \vec{\alpha}\vec{p} - \tau_1 \beta m_f + q\alpha_\mu B^\mu \right) \Phi_2(x) \end{aligned} \qquad (23)$$

It is still possible to write the Foldy-Wouthuysen transformation of the form

$$U_{FW} = U_{FW}^0 \left(1 + \delta_1 + \delta_2 + \delta_3 + ...\right) \quad (24)$$

$$U_{FW}^0 = \sqrt{\frac{E + \tau_3 \vec{\alpha}\vec{p}}{2E}} \left(1 \pm \frac{1}{E + \tau_3 \vec{\alpha}\vec{p}} \tau_3 \tau_1 \beta m \right) \quad (25)$$

The signs in (25) and (24), as well as in operators $C$ and $N$, are chosen depending on whether the transformation is applied to field $\Phi_1(x)$ (sign $+$), or it is applied to field $\Phi_2(x)$ (sign $-$).

Upon transformation (25), the form of the transformed functions (16), (17) and equations (13) remains unchanged. The free motion Hamiltonian density and the Hamiltonian density in the presence of interacting fields can be written in the form similar to (18), (22).

The $SU(2)$-invariant Standard Model with massive fermions can be derived with the isotopic space introduced. In case of interaction with gauge fields $B_\mu$, the Lagrangian with covariant derivative $D^\mu = \partial^\mu - iqB^\mu$ and fermion fields $\Phi_1 = \begin{pmatrix} \psi_R \\ \psi_L \end{pmatrix}, \Phi_2 = \begin{pmatrix} \psi_L \\ \psi_R \end{pmatrix}$ can be written as

$$\mathcal{L} = \bar{\Phi}_1 \gamma_\mu D^\mu \Phi_1 - \bar{\Phi}_1 \tau_1 m_f \Phi_1 + \bar{\Phi}_2 \gamma_\mu D^\mu \Phi_2 + \bar{\Phi}_2 \tau_1 m_f \Phi_2. \quad (26)$$

The Lagrangian allows obtaining the motion equations for fermion fields $\Phi_1, \Phi_2$ with fermion mass $m_f$ (see (23)).

$$p_0 \Phi_1(x) = \left(\vec{\alpha}\vec{p} + \tau_1 \beta m_f + q\alpha_\mu B^\mu \right) \Phi_1(x)$$
$$p_0 \Phi_2(x) = \left(\vec{\alpha}\vec{p} - \tau_1 \beta m_f + q\alpha_\mu B^\mu \right) \Phi_2(x)$$

Using the isotopic Foldy-Wouthuysen transformation (24), (25) one can obtain the $SU(2)$-invariant Hamiltonian density and motion equations for fermion fields with the sign in operators $U_{FW}, C, N, K_1, K_2, K_3...$ chosen properly (see (25)).

$$\mathcal{H}_{FW} = (\Phi_1)_{FW}^\dagger \left(\tau_3 E + qK_1(+m_f) + q^2 K_2(+m_f) + q^3 K_3(+m_f) + ...\right)(\Phi_1)_{FW} +$$

$$+ (\Phi_2)_{FW}^\dagger \left(\tau_3 E + qK_1(-m_f) + q^2 K_2(-m_f) + q^3 K_3(-m_f) + ...\right)(\Phi_2)_{FW} \quad (27)$$

$$p_0(\Phi_1)_{FW} = \left(\tau_3 E + qK_1(+m_f) + q^2 K_2(+m_f) + q^3 K_3(+m_f) + ...\right)(\Phi_1)_{FW}$$

$$p_0(\Phi_2)_{FW} = \left(\tau_3 E + qK_1(-m_f) + q^2 K_2(-m_f) + q^3 K_3(-m_f) + ...\right)(\Phi_2)_{FW} \quad (28)$$

In (27), (28) the notation $K_i(\pm m_f)$ indicates the sign of expressions containing fermion mass $m_f$.

Thus, with $SU(3) \times SU(2) \times U(1)$-invariance and each of its theoretical and experimental applications and consequences preserved, the Standard Model can be formulated without the requirement of interactions between Higgs bosons and fermions. Here, boson fields are responsible for the gauge invariance of the theory's boson sector only and interact only with gauge bosons $W_\mu^\pm, Z_\mu$, gluons and photons.

With such formulation of the theory, fermion masses are introduced from the outside. The theory has no vertices of Yukawa interactions between fermions and Higgs bosons and, therefore, there are no processes of scalar boson decay and generation of fermions $(H \to f\bar{f})$, no quarkonium states $\psi, \Upsilon, \theta$ including Higgs bosons, no interactions of Higgs bosons with gluons ($qqH$) and photons ($\gamma\gamma H$) via fermion loops, etc.

The suggested version of the Standard Model is, most likely, re-normalizable, because the boson sector of the theory remains massless till the introduction of the Higgs mechanism of spontaneous violation of the symmetry, and quantum electrodynamics with a massless quantum and massive electron and positron is the re-normalizable theory. Nevertheless, issues of re-normability of the suggested version of the Standard Model are to be studied profoundly.

Of course, the results of forthcoming experiments on searching for scalar bosons using the CERN's Large Hadron Collider would provide direct verification of the conclusions made in this paper concerning the formulation of the Standard Model without Higgs bosons' interactions with fermions.